\documentclass[showpacs,aps,prb,amsmath,amssymb]{revtex4}
\usepackage{graphicx}
\usepackage{bm}
\begin{document}
\renewcommand{\thefigure}{\arabic{figure}}
\renewcommand \thesection{\arabic{section}}
\renewcommand \thesubsection{\arabic{section}.\arabic{subsection}}
\baselineskip=0.7cm
\title{Single-particle and collective excitations in a charged Bose gas at finite temperature}
\author{B. Davoudi$^{1,2,3,*}$ and M. P. Tosi$^{1}$}
\address{$^1$ Classe di Scienze, Scuola Normale Superiore, Piazza dei Cavalieri 7, I-56126 Pisa, Italy\\
$^2$D\'epartment de Physique and Centre de Recherche en Physique du Solide, Universit\'e de Sherbrooke, Sherbrooke, Qu\'ebec, Canada J1K 2R1\\
$^3$Institute for Studies in Theoretical Physics and Mathematics, Tehran 19395-5531, Iran}
\begin{abstract}
	The main focus of this work is on the predictions made by the dielectric formalism in regard to the relationship between single-particle and collective excitation spectra in a gas of point-like charged bosons at finite temperature $T$ below the critical region of Bose-Einstein condensation. Illustrative numerical results at weak coupling ($r_s = 1$) are presented within the Random Phase Approximation. We show that within this approach the single-particle spectrum forms a continuum extending from the transverse to the longitudinal plasma mode frequency and leading to a double-peak structure as $T$ increases, whereas the density fluctuation spectrum consists of a single broadening peak. We also discuss the momentum distribution and the superfluidity of the gas.
\end{abstract}
\maketitle
\vspace{0.2 cm}

* Corresponding author, e-mail: b.davoudi@usherbrooke.ca

\section{Introduction}
\label{sec1}

The fluid of point-like charged bosons moving in a uniform neutralizing background (CBF or charged boson fluid) is a basic model in quantum statistical mechanics~\cite{1}, which is thought to be relevant to superconductors in the limit of small pairs\cite{2,3} and to dense plasmas of pressure-ionized helium in astrophysical objects\cite{4,5,6}. Interest in quantum Bose fluids with long-range interactions has recently extended to Bose-Einstein condensates of atoms with electromagnetically induced gravitational-like forces\cite{7} and to dipolar Bose-Einstein condensates\cite{8,9,10}.

The properties of the CBF at zero temperature have been extensively studied and are quite well known. The strength of the interactions is measured by the dimensionless parameter $r_s = (4\pi na_B^3/3)^{-1/3}$, where $n$ is the mean density and $a_B$ the Bohr radius, so that the weak-coupling regime corresponds to the limit of high density. This regime has been evaluated in early studies\cite{11,12,13} showing that the ground state is Bose-Einstein-condensed with a quantum depletion that increases with $r_s$. At very strong coupling ($r_s\approx 160$) the CBF undergoes Wigner crystallization\cite{14}, once the Coulomb repulsions become sufficiently strong to overcome the zero-point kinetic energy associated with localization of the bosons on lattice sites. The fluid state at intermediate and strong coupling has been evaluated by Quantum Monte Carlo techniques\cite{15} and has been studied by theoretical methods based on the use of Jastrow-Feenberg correlated wave functions (see Apaja {\it et al.}\cite{16} and references therein) or on self-consistent accounts of correlations (see Conti  {\it et al.}\cite{17} and references given there).

Relatively less is known on the behavior of the CBF with increasing temperature $T$ towards the critical temperature $T_c$ for the disappearance of the Bose-Einstein condensate. Hore and Frankel\cite{18} gave analytical results for the longitudinal dielectric response, and hence for the allowed modes of density oscillation and for the long-range screening of a static point charge, as functions of $T$ within the Random Phase approximation (RPA). This approach is expected to be valid at weak coupling and long wavelengths. Hansen {\it et al.}\cite{19} have drawn a schematic phase diagram and Bishop\cite{20} has evaluated the critical behavior near $T_c$ within a Hartree-Fock approach using RPA-screened interactions. Davoudi {\it et al.}\cite{21} have calculated the condensate fraction and the normal and anomalous momentum distributions within a Hartree-Fock-Bogoliubov (HFB) approach, adjusted to incorporate the Hugenholtz-Pines relation for the chemical potential.

There are in our view two main conceptual issues that are still worth studying in regard to the physical behavior of the CBF at finite $T$ below the critical region near $T_c$. The first concerns the role of the long-wavelength $f$-sum rules. In a normal Coulomb fluid the usual $f$-sum rule relates the first spectral moment of collective density fluctuations (or equivalently the spectral strength of the autocorrelation function of longitudinal current density fluctuations) to the total particle number density $n$ and sets the long-wavelength plasma frequency at the value $\omega_p=(4\pi n e^2/m)^{1/2}$, independently of temperature\cite{22}. In a superfluid, however, a further sum rule relates the superfluid density $n_s$ to the spectral strength of the autocorrelation function of the superfluid velocity at long wavelengths\cite{23}. At $T = 0$ one expects that $n_s = n$, but a normal component carrying entropy and viscosity appears and increases in density with increasing temperature. In the HFB approach\cite{21} the dispersion relation for quasi-particles was found to tend to the value $\omega_0=(4\pi n_0e^2/m)^{1/2}$ at long wavelengths, with $n_0$ being the condensate density, and this drops with increasing temperature up to vanishing at $T_c$. Even though the HFB approach is known to be unreliable in regard to conservation principles, it seems proper to further examine the role played at finite temperature by the two distinct particle conservation laws in the excitation spectra of a superfluid CBF consisting of a superfluid component and a normal component.

The second issue referred to above concerns the relationship between single-particle and collective density-wave excitations in the CBF at finite temperature. A theorem by Gavoret and Nozi\`{e}res\cite{24} states that in the presence of the condensate the two types of excitations in a fluid of neutral bosons at $T = 0$ have the same long-wavelength spectrum, apart from a factor. Chiofalo {\it et al.}\cite{25} have shown that this theorem also holds for the CBF at $T = 0$, once the Hugenholtz-Pines relation is satisfied. This implies, of course, that the single-particle excitation spectrum at long-wavelengths in a Bose-Einstein-condensed CBF at $T = 0$ has a gap equal to the plasma frequency - in contrast to a condensed neutral fluid where the Goldstone theorem requires the presence of a soft acoustic mode related to phase fluctuations. The above property of the CBF at $T = 0$ is again related to sum rules, as discussed by Chiofalo {\it et al.}\cite{25}. How this property evolves with increasing temperature still needs clarification.

In this work we examine the answers given to these questions by the so-called dielectric formalism\cite{26,27,28}, which was earlier developed for a neutral Bose gas and should be eminently suitable for a fluid with long-range interactions. This microscopic approach is based on a perturbative expansion and treats on the same footing the single-particle propagators and the density fluctuation propagator. The plan of the paper is briefly as follows. In Section 2 we recall the essential results of the perturbative expansion for the structure of the propagators, outline their relation to the linear response functions and to the momentum distribution of the fluid, and discuss poles and zeroes of these functions in the complex frequency plane. Section 3 reports illustrative numerical results that we have obtained within the RPA for the CBF at weak coupling ($r_s = 1$) and discusses the RPA predictions on its superfluidity and momentum distribution. Section 4  concludes the paper with a summary of its main results and a critique of the RPA.

\section{Dielectric formalism for a charged Bose gas}
\label{sec2}

We consider a three-dimensional homogeneous fluid of spinless Bose particles interacting through Coulomb forces described in Fourier transform by the potential $v_k=4\pi e^2/k^2$  for $k \neq 0$ and  $v_k= 0$ for $k = 0$. The latter condition takes care of charge neutrality as ensured by the presence of a uniform background at the same density $n$. 

The fluid is treated at finite temperature $T$ by the imaginary-time technique. The collective density-fluctuation correlations are described by the two-particle Green's function
\begin{equation}
\chi(k,\tau)=-\left\langle T_\tau\rho_{\bf k}(\tau)\rho_{-\bf k}(0)\right\rangle.
\end{equation}
The density fluctuation operator $\rho_{\bf k}$ is given as $\rho_{\bf k}=\sum_{\bf q}a^\dagger_{{\bf q}-{\bf k}/2}a_{{\bf q}+{\bf k}/2}$ in terms of the creation and destruction operators  $a^\dagger_{\bf k}$ and $a_{\bf k}$ for particles with momentum ${\bf k}$, which satisfy the usual bosonic commutation relations. The presence of a Bose-Einstein condensate made of $N_0$ particles induces hybridization between $\chi(k,\tau)$ and the single-particles Green's functions $G_\alpha^\beta(k,\tau)$, which are defined as
\begin{equation}
G_\alpha^\beta(k,\tau)=-\left\langle T_\tau b^\beta_{\bf k}(\tau)b^{\alpha\dagger}_{\bf k}(0)\right\rangle.
\end{equation}
Here $b^\alpha_{\bf k}=b_{\bf k}$ if $\alpha = +1$ and $b^\alpha_{\bf k}=b^\dagger_{-\bf k}$ if $\alpha = -1$, with $b_{\bf k}=a_{\bf k}-\sqrt{N_0}\delta_{\bf k}$.

We report below the main results of the diagrammatic analysis for the structure of the Green's functions in Fourier transform with respect to the imaginary-time variable, following the work of Sz\'{e}pfalusy and Kondor\cite{27}. The Green's function $G_\alpha^\beta(k,\tau)$ and its Fourier transform $G_\alpha^\beta(k,i\omega_n)$ are related by
\begin{equation}
G_\alpha^\beta(k,\tau)=k_BT\sum_n \exp{(i\omega_n\tau)}G_\alpha^\beta(k,i\omega_n)
\end{equation}
(and similarly for $\chi$), where $\omega_n=2\pi n k_B T$ are the Matsubara frequencies ($\hbar = 1$).

\subsection{Results of the perturbation expansion}

In the presence of the condensate every dynamical quantity can be separated into a ``singular'' reducible part (denoted by an index $s$) and a ``regular'' irreducible part (denoted by an index $r$). One can also extract the proper part of each quantity (indicated by a superscript \~{}) with respect to interaction lines. In particular, the density-density response function takes the form $\chi=\tilde \chi/(1-v_k \tilde \chi)$ where $\tilde \chi= \tilde \chi^{(s)}+\tilde \chi^{(r)}$. Thus the dielectric function $\varepsilon$, defined as usual through the relation  $1/\varepsilon=1+v_k\chi$ [22], is determined by the sum of two polarizability functions,
\begin{equation}
\varepsilon=1-v_k\tilde \chi^{(s)}-v_k\tilde \chi^{(r)}.
\end{equation}		
In this formalism the CBF is a two-component fluid, in which the singular component could be viewed as the condensate and the regular one as the noncondensate\cite{28} (see the RPA results in Section 3 below for an explicit approximate realization of this two-fluid model, in which the condensate density will turn out to be the same as the superfluid density). One can also introduce at this point a dielectric function  for $\varepsilon^{(r)}$ the regular component, $\varepsilon^{(r)}=1-v_k\chi^{(r)}$.

The structure of the single-particle Green's functions is similarly obtained with the help of the Dyson equations introducing the self-energies, $\Sigma_\alpha^\beta$. One finds
\begin{equation}
\label{gab}
G_{\alpha}^{\beta}=\frac{1}{D}\left[ \tilde N_{\alpha}^{\beta} +
\alpha\beta \tilde \Lambda_{-\alpha}\frac{v_k}{\epsilon^{(r)}} \tilde
\Lambda^{-\beta} \right]
\end{equation}
where $\tilde\Lambda_\alpha$ are the vertex functions,  $\tilde N_{\alpha}^{\beta}=\delta_{\alpha}^{\beta}(\alpha i\omega_n +\varepsilon_k)$  with $\varepsilon_k=k^2/(2m)$ referred to the chemical potential $\mu$, and
\begin{equation}
D =(i\omega_n  -\varepsilon_{\bf k}-\Sigma_+^+)(i\omega_n  +\varepsilon_{\bf k}+\Sigma_-^-)+ \Sigma_+^- \Sigma_-^+.
\end{equation}
The proper self-energies $\tilde\Sigma_\alpha^\beta$ are introduced by setting $\Sigma_\alpha^\beta=\tilde \Sigma_\alpha^\beta+\tilde \Lambda_\alpha v_k \tilde \Lambda^\beta/\epsilon^{(r)}$, and the proper single-particle Green's functions are given by $ \tilde G_\alpha^\beta = \tilde N_\alpha^\beta/\tilde D$ where $\tilde D$ is given by Eq.~(6) with $\Sigma_\alpha^\beta$ replaced by $\tilde\Sigma_\alpha^\beta$. Finally, the vertex functions relate the proper singular susceptibility $\tilde\chi^{(s)}$ to the proper self-energies through $\tilde \chi^{(s)}=\tilde \Lambda_{\alpha} \tilde G_{\alpha}^{\beta} \tilde \Lambda^{\beta}$  (here and in the following summation over repeated indices is implicit). 

 The presence of the condensate thus induces a close relationship between charge density fluctuations and single-particle excitations. Using the equality
\begin{equation}
\label{deed}
\tilde D\epsilon=D\epsilon^{(r)} 
\end{equation}
the expressions for $\chi$ and $G_\alpha^\beta$ can be written in the form
\begin{equation}
\chi=\frac{\tilde \chi^{(r)}}{\varepsilon} +\frac{\tilde \Lambda_\alpha\tilde N_\alpha^\beta  \tilde \Lambda^\beta}{D\epsilon^{(r)}}
\label{chif}
\end{equation}
and
\begin{equation}
\label{gabf}
G_\alpha^\beta=\frac{\tilde N_{\alpha}^{\beta}}{D} + \alpha \beta\frac{ \tilde \Lambda_{-\alpha} v_k  \tilde 
\Lambda^{-\beta}}{\tilde D\epsilon},
\end{equation}
where
\begin{equation}
\varepsilon=\varepsilon^{(r)}+v_k\Lambda_\alpha\tilde N_\alpha^\beta  \tilde \Lambda^\beta/\tilde D
\end{equation}
and
\begin{equation}
D=\tilde D+v_k\Lambda_\alpha\tilde N_\alpha^\beta  \tilde \Lambda^\beta/\varepsilon^{(r)}.
\end{equation}
These expressions will be used later below to discuss the poles and zeroes of the various response functions. It may also be explicitly remarked that the vertex functions vanish at $T_c$, so that the single-particle and density-fluctuation spectra become uncoupled for $T \geq T_c$.

\subsection{Superfluidity and momentum distribution}

We pause at this point to set out the expressions for some further physical properties of the Bose fluid that can be obtained from knowledge of its Green's functions. The autocorrelation function  $\chi_{v_sv_s}$ of the superfluid velocity $v_s$ is obtained from the $G^{(0)}$ component of the matrix of single-particle Green's functions on the $\tau^{(0)}$ Pauli matrix (the $2\times 2$ identity matrix),
\begin{equation}
\chi_{v_sv_s}=-\frac{k^2}{m^2 n_0}\,G^{(0)}.
\end{equation}
This is related to the autocorrelation function $\chi_{\phi\phi}$ of phase fluctuations as $\chi_{v_sv_s}=k^2\chi_{\phi\phi}/m^2$. The superfluid mass density $\rho_s=mn_s$ follows from $\chi_{v_sv_s}$ by means of the sum rule\cite{23}
\begin{equation}
\frac{1}{\rho_s}=\lim_{k\rightarrow 0}\chi_{v_sv_s}(k,0)= \lim_{k\rightarrow 0}\int_{-\infty}^{+\infty}\frac{d\omega}{\pi}\frac{{\rm Im} \chi_{v_sv_s}(k,\omega)}{\omega}.
\end{equation}
The last identity in Eq.~(13) emphasizes the meaning of this relation as an $f$-sum rule on the superfluid component at each temperature $T$, on account of the expression ${\bf J}_s=n_s{\bf v}_s$ for the superfluid current density ${\bf J}_s$ and by comparison with the $f$-sum rule on the total density $n$,
\begin{equation}
n=\int_{-\infty}^{\infty}\frac{d\omega}{\pi}\frac{{\rm Im} \chi_{JJ}(k,\omega)}{\omega}.
\end{equation}
Here $\chi_{JJ}$ is the longitudinal current-current response function. Of course, the continuity equation for density fluctuations relates $\chi_{JJ}$ to the density-density response function $\chi$.

Equation~(13) leads to the Josephson relation for the superfluid density,
\begin{equation}
\frac{n_s}{n_0}=\lim_{k\rightarrow 0}\left[\frac{k^2/(2m) -\mu +\Sigma_+^+-\Sigma_+^-}{k^2/(2m)}\right].
\end{equation}
An alternative definition of the superfluid mass density is obtained\cite{29} by writing it as  $\rho_s=mn-\rho_n$ , where the mass density $\rho_n$ of the normal component is determined by the autocorrelation function of the transverse current density $\chi_T$ as
\begin{equation}
\rho_n=\lim_{k\rightarrow 0}\chi_{T}(k,0).
\end{equation}
As discussed in the book of Forster\cite{30}, a proof of the equivalence between the definition of superfluid density that is given in Eq.~(15) from properties of the single-particle excitation spectrum and the definition that follows from Eq.~(16) in terms of the transverse current correlations is still a question of continued interest.

Turning to the one-body momentum distribution function $n(k)$ of the Bose fluid, this is related to the single-particle Green's function $G_+^+$ by
\begin{equation}
n(k) = -k_BT\sum_n \exp(i\omega_n\eta)G_+^+(k,i\omega_n)
\end{equation}
where $\eta$ is a positive infinitesimal. The summation can be written in the form of a contour integral by means of the relation
\begin{equation}
\label{oint}
\sum_n  \exp(i\omega_n\eta)G_+^+(k,i\omega_n)=\frac{1}{2\pi ik_BT}\oint_{c_1}dz\frac{\exp(\eta z)}{\exp(z/k_BT)-1} G_+^+(k,z).
\end{equation} 
as is depicted in Fig.~1. Equation~(18) follows from the fact that the poles of $[\exp(z/k_BT)-1]^{-1}$ lie at the points $i\omega_n$ on the imaginary axis in Fig.~1. The contour of integration can be conveniently deformed from $c_1$ to $c_2$ under conditions such that $G_+^+(k,z)$ only has single poles with a small imaginary part (see Fig.~1 again). In particular this is true at $T= 0$ for the CBF at long wavelengths, where the poles lie at the plasma frequency $\omega_p=(4\pi n e^2/m)^{1/2}$ and infinitesimally away from the real axis\cite{25}. In this case we find
\begin{equation}
\lim_{k\rightarrow 0} n(k)=\left.\frac{v_k \tilde \Lambda ^2}{2 \omega_p \varepsilon^{(r)}}\right|_{k\rightarrow 0}
\end{equation}
and comparison with the exact result
\begin{equation}
\lim_{k\rightarrow 0} n(k)=\frac{m n_0 \omega_p}{2 k^2}
\end{equation}
obtained by Chiofalo {\it et al.}\cite{25} yields
\begin{equation}
\lim_{k\rightarrow 0} \tilde \Lambda^2(k,\omega_p)/\epsilon^{(r)}(k,\omega_p)=n_0
\end{equation}
at $T = 0$. A similar approach can be used to infer that in a neutral Bose gas the momentum distribution scales as $k^{-1}$ at low momenta.

\subsection{Spectral structures}

We return to Eq.~(7) and notice that, if the dielectric function $\varepsilon$ of the fluid vanishes at some frequency ($z_1$, say) in the complex frequency plane, the function $D$ vanishes at the same frequency. Indeed, from Eq.~(10) $\varepsilon^{(r)}$ will not vanish there as long as the vertex functions are different from zero. A ``ghost'' of the collective oscillations of the particle density, which are determined by the zeroes of $\varepsilon$\cite{22}, will therefore be present in the single-particle excitation spectrum below $T_c$. In particular, in the long wavelength limit at $T = 0$ both spectra are delta-function peaks lying at the plasma frequency $\omega_p$ and the Gavoret-Nozi\`eres theorem holds, as demonstrated by Chiofalo {\it et al.}\cite{25}. The hybridization of the poles in $\chi$ and in the $G$'s can be understood by viewing the single-particle spectrum as associated to excitation of a particle out of the condensate, which should be driven to oscillate by the Coulomb forces and in turn drive an oscillation of the whole fluid at the plasma frequency.

However, the two types of spectra should be expected to become substantially different at $T>0$, when a normal (non-superfluid) component of the fluid emerges and permits different damping mechanisms for density fluctuations and for single-particle excitations through the dissipation of momentum and energy. Damping by excitation of longitudinal currents is operative for the former, and by excitation of both longitudinal and transverse currents for the latter. Transverse current fluctuations correspond to a pole in the dielectric function\cite{22} at some frequency ($z_2$, say) in the complex frequency plane. From Eqs.~(7) and (10) this can correspond to a zero in $\tilde D$ and$/$or to a pole in $\varepsilon^{(r)}$ at the same frequency. One may thus envisage the spectrum of single-particle excitations at finite temperature as forming a continuum that extends from a lower edge determined by the onset of transverse currents and is surmounted in its upper part by a ghost due to the coupling to collective density fluctuations. The latter, on the other hand, correspond to longitudinal current fluctuations and are not coupled to transverse current fluctuations. We shall explicitly see in Section 3 below how these expectations are confirmed by the RPA calculations.

One last point to be made concerns the dielectric function $\varepsilon^{(r)}$, which gives the dielectric response of the ``regular'' component of the fluid. From Eq.~(11) we see that a zero of this function at some frequency ($z_3$, say) in the complex frequency plane corresponds to a pole in $D$ at the same frequency. It is tempting to view this fact as expressing the effect of transverse single-particle motions in driving the regular component. It should also be noted that the form of $\varepsilon^{(r)}(k,z)$ may depend on the collisionality of the gas, which may induce a dependence on $kl$ where $l$ is the mean free path between collisions (see the discussion given by Pippard\cite{31} for an analog in the gas of conduction electrons in metals). This may even affect the ability of the regular component to screen the density fluctuations and the single-particle motions.

We proceed from this heuristic overview to examine how the expected spectral features are brought out in the approximate theoretical model afforded by the RPA.

\section{The Random Phase Approximation}
\label{sec3}

As developed by Hore and Frankel\cite{18} and by Sz\'{e}pfalusy and Kondor\cite{27}, the RPA for a Bose fluid at temperature $T < T_c$ adopts the ideal Bose gas as reference and accounts for the interactions by a mean-field treatment neglecting all short-range correlations. In a fluid of charged particles these approximations are expected to be valid at long wavelengths ($k \rightarrow 0$) and for low values of the Coulomb coupling strength $r_s$\cite{22}.

As a first step, therefore, the chemical potential is set to zero and the proper susceptibilities are taken to be those of the ideal Bose gas, that is
\begin{equation}
\tilde \chi^{(s)}(k,\omega)=\frac{2 n_0
\varepsilon_k}{\omega^2-\varepsilon_k^2}
\end{equation}
for the condensate and
\begin{equation}
\tilde\chi^{(r)}(k,\omega)=\sum_{{\bf q}\neq 0,- {\bf k}}\frac{f({\bf q})-f({\bf q}+{\bf k})}{\omega-(\varepsilon_{{\bf q}+{\bf
k}}-\varepsilon_{\bf q})}
\end{equation}
for the thermal cloud, with $f({\bf q})=[\exp(\varepsilon_q/k_BT)-1]^{-1}$ being the statistical distribution function of the ideal Bose gas below $T_c$ (we are for simplicity omitting to write an infinitesimal imaginary part in the denominators). The function $\varepsilon^{(r)}(k,\omega)$ accordingly describes the dielectric response of the non-interacting thermal cloud and keeps no account of interparticle collisions: in the language of quantum hydrodynamics, the thermal cloud is taken to be in the collisionless regime.

At the next step one sets $\tilde\Lambda_\alpha=\tilde\Lambda^\alpha=\sqrt{n_0}$ and $\tilde\Sigma_\alpha^\beta=0$. One accordingly finds
\begin{equation}
\tilde D(k,\omega)=\omega^2-\varepsilon_k^2
\end{equation}
and
\begin{equation}
\Sigma_\alpha^\beta(k,\omega)=\frac{n_0v_k}{\varepsilon^{(r)}(k,\omega)}.
\end{equation}
The approximations listed above finally yield
\begin{equation}
\varepsilon(k,\omega)=\varepsilon^{(r)}(k,\omega)-\frac{2n_0v_k\varepsilon_k}{\omega^2-\varepsilon_k^2},
\end{equation}
\begin{equation}
D(k,\omega)=\omega^2-\varepsilon_k^2-\frac{2n_0v_k\varepsilon_k}{\varepsilon^{(r)}(k,\omega)},
\end{equation}
and
\begin{equation}
G_+^+(k,\omega)=\frac{\omega+\varepsilon_k+n_0v_k/\varepsilon^{(r)}(k,\omega)}{D(k,\omega)}.
\end{equation}
In the RPA the self-energies are determined by Coulomb interactions with the condensate screened by the ideal thermal cloud and the transverse current modes lie at frequency $\varepsilon_k$.

The Bogoliubov approximation is recovered by setting  $\varepsilon^{(r)}(k,\omega)= 1$. As discussed by Sz\'{e}pfalusy and Kondor\cite{27}, such a Bogoliubov regime can be met in a neutral Bose gas under somewhat special conditions at intermediate temperatures.

\subsection{Excitation spectra}

It is immediately seen from Eqs.~(26) and (27) that in the RPA the single-particle Green's functions and the density-density response function have a pole at the same frequency, which is determined by the zero of $\varepsilon(k,z)$ in the complex frequency plane. In addition the single-particle Green's functions may have a second pole determined by a pole in $\varepsilon^{(r)}(k,z)$.

The calculation of the susceptibility of the ideal Bose gas\cite{18} yields
\begin{equation}
{\rm Re}\tilde \chi^{(r)}(k,z)=\frac{m}{4\pi^2k}\int_0^\infty dq\,q\,n(q)\ln{\left|\frac{m^2z^2-(qk-\frac{1}{2}k^2)^2}{m^2z^2-(qk+\frac{1}{2}k^2)^2}\right|}
\end{equation}
and
\begin{equation}
{\rm Im}\tilde\chi^{(r)}(k,z)=\frac{m}{4\pi k}
\ln\left|\frac{1-\exp[-(z-\varepsilon_k)^2/(4k_BT\varepsilon_k)]}
{1-\exp[-(z+\varepsilon_k)^2/(4k_BT\varepsilon_k)]}\right|.
\end{equation}
In the limit $k\lambda_{dB}\gg 1$, where $\lambda_{dB}=(1/2mk_BT)^{1/2}$ is the thermal de Broglie wavelength,  $\tilde\chi^{(r)}$ acquires the single-pole structure
\begin{equation}
\tilde\chi^{(r)}(k,z)\rightarrow \frac{2 (n-n_0)
\varepsilon_{k}}{z^2-\varepsilon_{k}^2}.
\end{equation}
Under conditions such that the imaginary part of $\tilde\chi^{(r)}$ can be neglected, the zero of $\varepsilon(k,\omega)$ yields the dispersion relation
\begin{equation}
\label{disp}
E_k=\sqrt{\varepsilon_k^2+ 2 n_0 v_k \varepsilon_k  
/\epsilon^{(r)}(k,E_k)}.
\end{equation}
and in particular, in the extreme low-temperature limit where Eq.~(31) holds at all wave numbers, one finds $E_k\rightarrow\sqrt{ 2 n v_k \varepsilon_k +\varepsilon_k^2+}\rightarrow \omega_p$ at long wavelengths.

The dispersion relation obtained from Eq.~(32) in the RPA for $r_s = 1$ is shown in Fig.~2 at increasing values of the temperature (in units of $\hbar^2/(k_Bma_B^2)$, such that $T_c = 1.27$). A $k^2$ term is present at finite $T$, where one finds\cite{18}
\begin{equation}
E_k^2=\omega_p^2+A(n-n_0)k_BT\varepsilon_k+\varepsilon_k^2
\end{equation}
with a numerical coefficient $A=3\zeta(\frac{5}{2})/\zeta(\frac{3}{2})$ given in terms of Riemann zeta functions. The effect of short-range correlations beyond the RPA is to introduce a term of order $k^2$ even at $T = 0$ with a negative coefficient\cite{15}, which is responsible for the presence of a roton minimum in the dispersion relation.

Figure~3 shows the spectrum of collective density oscillations at a relatively small value of the wave number ($k = 1/a_B$), as calculated in the RPA for the weakly coupled CBF ($r_s = 1$) at increasing values of temperature. The spectrum of single-particle excitations is shown in Fig.~4 at the same wave number and temperatures. At such wave number the plasmon excitation is still seen in Fig.~3 as a sharp low-temperature peak, which with increasing $T$ moves upwards as predicted by Eq.~(33) and at the same time broadens. Figure~4 shows that the spectrum of single-particle excitations is very substantially different: its main peak is definitely broader at the lowest temperature, and as $T$ rises this peak broadens much more rapidly and at the same time acquires a low-frequency plateau which at high temperature is developing into a second peak. The sharp feature at the bottom edge of the spectrum lies for all values of $T$ exactly at the frequency $\varepsilon_k$ of transverse current excitations, which in the units used in Fig.~4 is  $\varepsilon_k= 0.5$. The shift of oscillator strength from the higher to the lower peak in the single-particle spectrum with increasing $T$ signals the progressive reduction of the condensate and hence of the hybridization with collective density fluctuations hybridization vanishes with the condensate at $T_c$.

\subsection{Superfluidity}

From Eq.~(12) we find that the autocorrelation function of the superfluid velocity within the RPA takes the form
\begin{equation}
\chi_{v_sv_s}(k,\omega)=-\frac{k^2}{m^2}\frac{v_k/\epsilon^{(r)}(k,\omega)}{\omega^2-\varepsilon_k^2-2n_0v_k \varepsilon_k/\epsilon^{(r)}(k,\omega)}.
\end{equation}
The sum rule in Eq.~(13) (or equivalently the Josephson relation in Eq.~(15)) then yields
\begin{equation}
\rho_s=mn_0,
\end{equation}
that is, within the RPA the superfluid density and the condensate density coincide.

The same result is found from the second definition of the superfluid density based on Eq.~(16) for the mass density of the normal component. From the dielectric formalism we have $\chi_T=\tilde \chi_T$ and within the RPA the latter function is approximated by the transverse current response of the ideal Bose gas:
\begin{equation}
\label{xt}
\tilde\chi_T({\bf k},\omega)=\frac{1}{m^2}\sum_{i,j}\left(\hat{k}_i\hat{k}_j-\delta_{ij}
\right) \sum_{{\mathbf q}\neq 0, -{\mathbf k}} \left({\bf
q}+\frac{1}{2}{\bf k}\right)_i\left({\bf
q}+\frac{1}{2}{\bf k}\right)_j \frac{f({\bf q})-f({\bf q}+{\bf
k})}{\omega-(\varepsilon_{{\bf q}+{\bf
k}}-\varepsilon_{\bf q})}.
\end{equation}
The imaginary part of the expression in Eq.~(36) can be evaluated analytically by the method used for the ideal Fermi gas by Nifos\`{i} {\it et al.}\cite{32}. We set ${\rm Im}\tilde\chi_T(k,\omega)=A_{\bf k}(\omega)-A_{\bf k}(-\omega)$, where
\begin{equation}
A_{\bf  k}(\omega)=\frac{\pi}{m^2} \sum_{\bf q} \frac{({\bf k}\cdot
{\bf q})^2}{k^2} f({\bf q}) \delta(\omega-\varepsilon_{{\bf q}+{\bf k}}+\varepsilon_{\bf
q}),
\end{equation}
By direct integration we obtain
\begin{equation}
A_{\bf  k}(\omega)=-\frac{m (k_BT)^2}{2\pi}\,{\rm PolyLog}\left[2,
\exp\left(\frac{(\omega-\varepsilon_k)^2}{4 k_BT\varepsilon_k }-\frac{\mu}{k_BT}\right)\right]
\end{equation}
where PolyLog$[n,z]$ denotes the n$^{\rm th}$-order polylogarithm function of the variable $z$. Numerical integration yields $\tilde\chi_T(k,0)$ and hence with Eq.~(16) we find $\rho_n=m(n-n_0)$ in the RPA. These results extend to finite temperature the findings of Alexandrov and Beere\cite{33}, who have calculated the transverse response using the Bogoliubov theory.

\subsection{Momentum distribution}

The momentum distribution in the RPA is obtained from Eqs.~(17) and (18) using the result in Eq.~(28) for the Green's function $G_+^+$. It is useful in performing the integration in the complex frequency plane to rewrite Eq.~(28) in the form
\begin{equation}\label{gff}
G_+^+(k,z)=\frac{g_+(k,z)}{f(k,z)-E_k^B}-\frac{g_-(k,z)}{f(k,z)+E_k^B}
\end{equation} 
where $f(k,z)=\sqrt{z^2+2n_0v_k\varepsilon_k[1-1/\varepsilon^{(r)}(k,z)]}$, $E_k^B=\sqrt{2n_0\varepsilon_kv(k)+\varepsilon_k^2}$  is the Bogoliubov dispersion relation, and
\begin{equation}
g_{\pm}(k,z)=\frac{\varepsilon_k+n_0 v_k/\epsilon^{(r)}(k,z)}{2 E_k^B}\pm\frac{z}{2f(k,z)}.
\end{equation}
The integration in Eq.~(18) can be carried out by deforming the contour from $c_1$ to $c_2$ as depicted in Fig.~1, provided that the spectral broadening due to thermal excitations can be neglected. The result is
\begin{equation}
\label{nff}
n(k)=\frac{1}{|\partial f/\partial z|_{z=E_k}}\left[\frac{g_+(k,E_k)}{\exp(E_k/k_BT)-1} +\frac{g_-(k,E_k)}{1-\exp(-E_k/k_BT)}\right],
\end{equation}
where $E_k$ is the dispersion relation given in Eq.~(33). A plot of Eq.~(41) is shown in Fig.~5. 

It is seen from Fig.~5 that Eq.~(41) misses the subleading term in the momentum distribution at low momenta, which is known to be proportional to $1/k$ with a positive coefficient\cite{15} and is present in the HFB approach\cite{21}. It may also be recalled that the HFB calculations yield a sizable quantum depletion of the condensate in the CBF already at $r_s = 1$, as well as a sizable upward shift of the critical temperature over the ideal-gas value\cite{21}. In principle these effects could be examined in an extended RPA approach, where a measure of self-consistency may be included through a redetermination of the condensate fraction from the one-body momentum distribution function. Such an effort seems, however, hardly worthwhile.

\section{Concluding remarks}
\label{sec4}

In conclusion, we have studied the dynamical properties of a weakly coupled charged Bose gas at finite temperature by means of the dielectric formalism. The formalism allows one to treat at the same time and on the same footing both the collective density fluctuations and the single-particle motions, keeping account of the hybridization between one-body and two-body Green's functions typical of a Bose-Einstein-condensed fluid. The formalism could also be extended to treat spinorial Bose gases.

We have illustrated the dielectric formalism by numerical calculations carried out within the random phase approximation, in which the dynamical effects of the noncondensate enter through the screening of the Coulomb interactions. We have seen that this approximation yields a qualitatively correct view of the spectra of collective and single particle motions, in which the latter at finite temperature are directly coupled to both the longitudinal and the transverse current density fluctuations. The spectral strength of the single-particle motions gradually shifts from the former to the latter as $T$ is raised towards $T_c$. It would indeed be absurd that, as $T$ is near to entering the critical region, the excitation of a single particle in the presence of a disappearing condensate should still be expected to require that the whole gaseous cloud be set into oscillation at the plasma frequency.

The main limitation of the dielectric approach is that the quantum fluid is treated as composed of a condensate and a noncondensate, rather than as consisting of a superfluid and a normal fluid, so that the superfluid fraction is an outcome of the calculation rather than its basic input. This viewpoint is, of course, intrinsic to the Bogoliubov-Hugenholtz-Pines theoretical framework. In the RPA the superfluid fraction turns out to be the same as the condensate fraction, and the shortcomings of the one-body momentum distribution that it predicts have been pointed out in the body of the paper. An account of this and other quantitative defects of the RPA predictions, such as the lack of a roton minimum in the dispersion curve of collective excitations, would require a refined treatment of the proper self-energies and the inclusion of short-range correlations between density fluctuations. The so-called one-loop approximation\cite{28} could be useful in these respects. Further spectral structures may also be expected to arise\cite{26}. 

\section{Acknowledgements}
\label{sec5}

	We acknowledge the early contributions to this work by Dr. A. Minguzzi and the support received through the PRIN2003 Program of MIUR and through an Advanced Research Initiative of SNS. One of us (MPT) thanks Dr. V. E. Kravtsov and the Condensed Matter Theory Group of the Abdus Salam ICTP for their hospitality during the final stages of this work.

\newpage

\begin{figure}
\begin{center}
\includegraphics[scale=0.5]{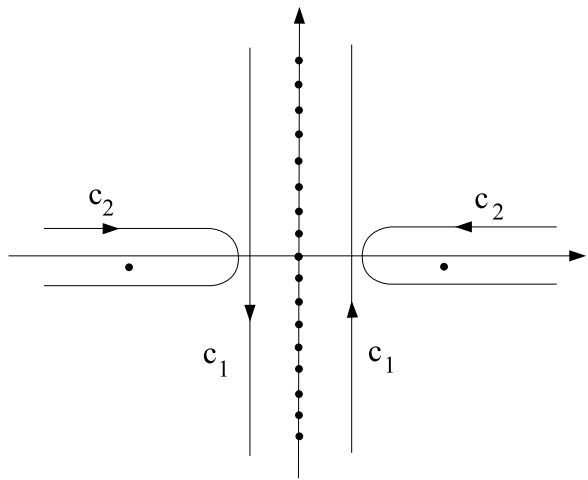}
\caption{Integration contours in the complex frequency plane.}
\end{center}
\end{figure}

\begin{figure}
\begin{center}
\includegraphics[scale=0.5]{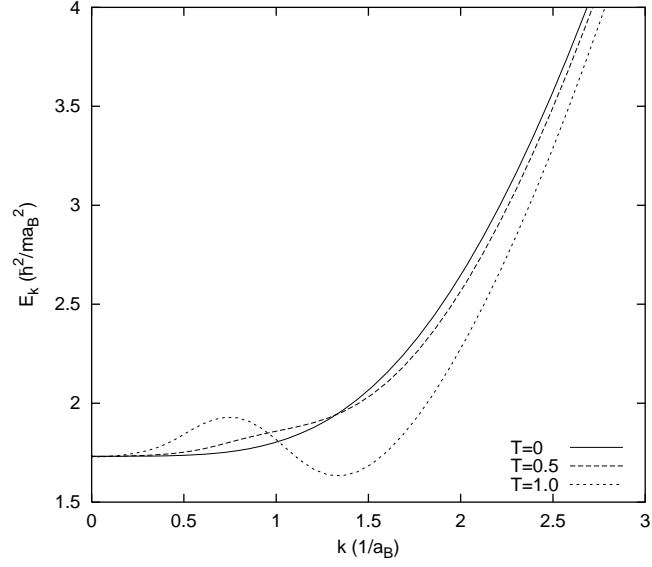}
\caption{RPA dispersion relation $E_k$ (in units of $\hbar^2/(ma_B^2)$) as a function of wave number $k$ (in units of $1/a_B$) for the CBF at $r_s = 1$,  at the values of the temperature $T$ indicated in the figure (in units of $\hbar^2/(k_Bma_B^2)$ - in these units the critical temperature for Bose-Einstein condensation of the ideal Bose gas is $T_c = 1.27$).}
\end{center}
\end{figure}

\begin{figure}
\begin{center}
\includegraphics[scale=0.5]{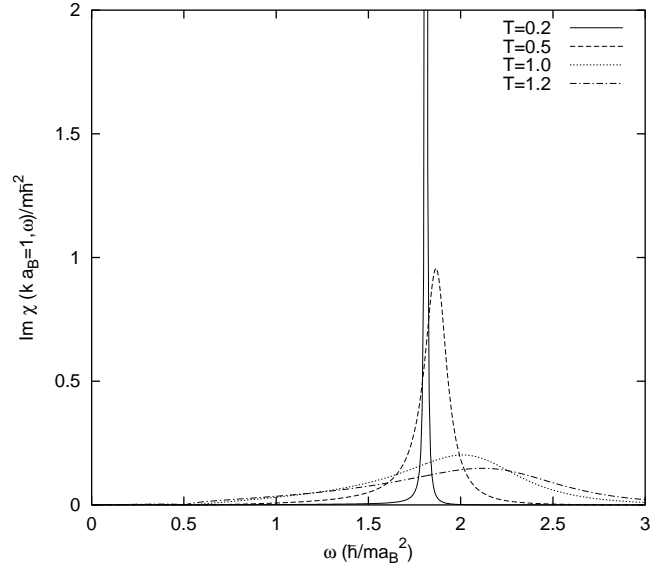}
\caption{Spectrum of collective density excitations at $ka_B = 1$ for the CBF at $r_s = 1$ in the RPA,  at the values of the temperature $T$ indicated in the figure (in units of $\hbar^2/(k_Bma_B^2)$).}
\end{center}
\end{figure}

\begin{figure}
\begin{center}
\includegraphics[scale=0.5]{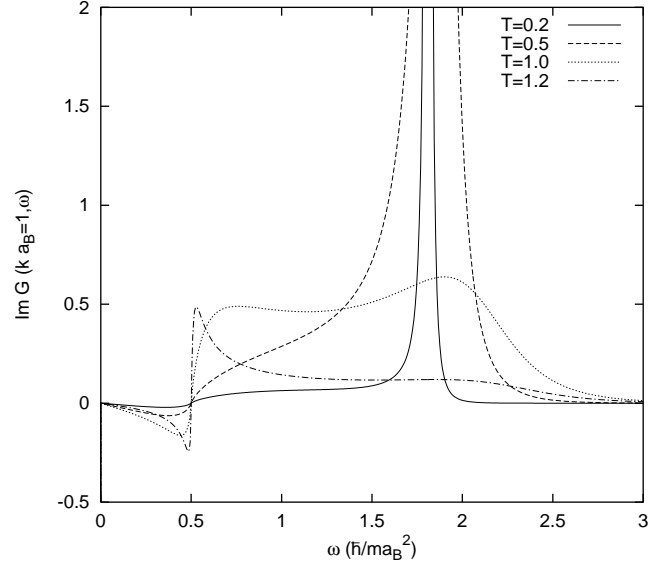}
\caption{Spectrum of single particle excitations at $ka_B = 1$ for the CBF at $r_s = 1$ in the RPA,  at the values of the temperature $T$ indicated in the figure (in units of $\hbar^2/(k_Bma_B^2)$).}
\end{center}
\end{figure}

\begin{figure}
\begin{center}
\includegraphics[scale=0.5]{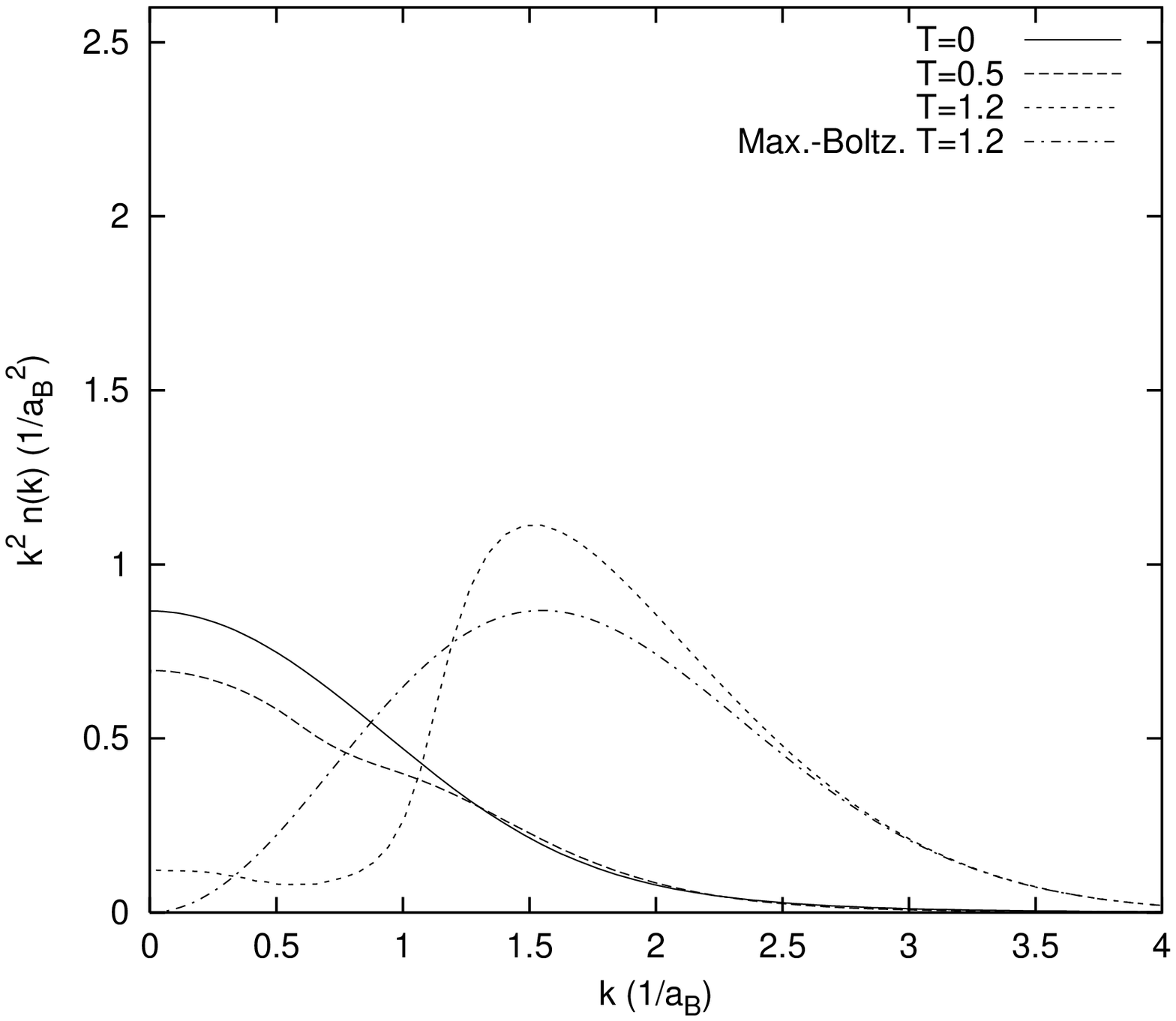}
\caption{Momentum distribution $k^2n(k)$ as a function of wave number $k$ (in units of $1/a_B$) for the CBF at $r_s = 1$ in the RPA,  at the values of the temperature $T$ indicated in the figure (in units of $\hbar^2/(k_Bma_B^2)$). The classical Maxwell-Boltzmann distribution is also shown.}
\end{center}
\end{figure}

\begin{thebibliography}{99}
\bibitem {1}	F. Y. Wu, E. Feenberg, Phys. Rev. {\bf 128}, 943 (1962).
\bibitem{2}	M. R. Schafroth, Phys. Rev. {\bf 96}, 1954 (1149).
\bibitem{3}	A. S. Alexandrov, N. F. Mott, Supercond. Sci. Technol. {\bf 6}, 215 (1993).
\bibitem{4}	B. W. Ninham, Phys. Lett. {\bf 4}, 278 (1963).
\bibitem{5}	S. Schramm, K. Langange, S. E. Koonin, Astrophys. J. {\bf 397}, 579 (1882).
\bibitem{6}	H. -M. M¸ller, K. Langange, Phys. Rev. C {\bf 49}, 524 (1996).
\bibitem{7}	D. O'Dell, S. Giovanazzi, G. Kurizki, V. M. Akulin, Phys. Rev. Lett. {\bf 84}, 5687 (2000).
\bibitem{8}	L. Santos, G. V. Shlyapnikov, M. Lewenstein, Phys. Rev. Lett. {\bf 90}, 250403 (2003).
\bibitem{9}	D. H. J. O'Dell, S. Giovanazzi, G. Kurizki, Phys. Rev. Lett. {\bf 90}, 110402 (2003).
\bibitem{10}	A. Minguzzi, N, H. March, M. P. Tosi, Phys. Rev. A {\bf 70}, 025601 (2004).
\bibitem{11}	L. L. Foldy, Phys. Rev. {\bf 124}, 649 (1961).
\bibitem{12}	K. A. Br¸ckner, Phys. Rev. {\bf 156}, 204 (1967).
\bibitem{13}	S. K. Ma and C. W. Woo, Phys. Rev. {\bf 159}, 165 (1967).
\bibitem{14}	D. M. Ceperley, B. J. Alder, Phys. Rev. Lett. {\bf 45}, 566 (1980).
\bibitem{15}	S. Moroni, S. Conti, M. P. Tosi, Phys. Rev. B {\bf 53}, 9688 (1996).
\bibitem{16}	V. Apaja, J. Halinen, V. Halonen, E. Krotscheck, M. Saarela, Phys. Rev. B {\bf 55}, 12925 (1997).
\bibitem{17}	S. Conti, M. L. Chiofalo, M. P. Tosi, J. Phys.: Condens. Matter {\bf 6}, 8795 (1994).
\bibitem{18}	S. R. Hore and N. E. Frankel, Phys. Rev. B {\bf 12}, 2619 (1975).
\bibitem{19}	J. P. Hansen, B. Jancovici, D. Schiff, Phys. Rev. Lett. {\bf 29}, 991 (1972).
\bibitem{20}	R. F. Bishop, J. Low Temp. Phys. {\bf 15}, 601 (1974).
\bibitem{21}	B. Davoudi, A. Minguzzi, M. P. Tosi, Phys. Rev. B {\bf 65}, 144507 (2002).
\bibitem{22}	D. Pines, P. Nozi\`{e}res, {\it The Theory of Quantum Liquids} (Benjamin, New York, 1966).
\bibitem{23}	P. C. Hohenberg, P. C. Martin, Ann. Phys. (NY) {\bf 34}, 291 (1965).
\bibitem{24}	J. Gavoret, P. NoziËres, Ann. Phys. (NY) {\bf 28}, 349 (1964).
\bibitem{25}	M. L. Chiofalo, S. Conti, M. P. Tosi, J. Phys.: Condens. Matter {\bf 8}, 1921 (1996).
\bibitem{26}  A. Griffin and T. H. Cheung, Phys. Rev. A {\bf 7}, 2086 (1973). 
\bibitem{27}	P. SzÈpfalusy, I. Kondor, Ann. Phys. (NY) {\bf 82}, 1 (1974).
\bibitem{28}	A. Griffin, {\it Excitations in a Bose-condensed Liquid}, Chapter 5. Cambridge University Press, Cambridge, 1993.
\bibitem{29}	G. Baym, in {\it Mathematical Methods in Solid State and Superfluid Theory} (Oliver and Boyd, Edinburgh, 1969), Eds. R. C. Clark and G. H. Derrick, p. 121. 
\bibitem{30}	D. Forster, {\it Hydrodynamic Fluctuations, Broken Symmetry, and Correlation Functions} (Benjamin, New York, 1975).
\bibitem{31}  A. B. Pippard, Rep. Progr. Phys. {\bf 23}, 176 (1960).
\bibitem{32}	R. NifosÏ, S. Conti, M. P. Tosi, Phys. Rev. B {\bf 58}, 12758 (1998).
\bibitem{33}	A. S. Alexandrov and W. H. Beere, Phys. Rev. B {\bf 51}, 5887 (1995).
\end{thebibliography}
\end{document}